\documentstyle[12pt,amssymbols]{article}
\input math_macros.tex

\begin{document}
\begin{titlepage}
\begin{center}
October 1994          \hfill LBL-37559 \\
          \hfill    UCB-PTH-95/27 \\
         \hfill   hep-th/9507155 \\

\vskip .25in

{\large \bf Convergent sequences of perturbative approximations for the
anharmonic oscillator\\
I. Harmonic approach}\footnote{This work was supported in part
by the Director, Office of
Energy Research, Office of High Energy and Nuclear Physics, Division of
High Energy Physics of the U.S. Department of Energy under Contract
DE-AC03-76SF00098 and in part by the National Science Foundation under
grant PHY-90-21139.}

\vskip .25in

B. Bellet \\
P. Garcia\\

{\em Laboratoire de Physique Math\'ematique\\
Universit\'e de Montpellier II-CNRS\\
34095 Montpellier c\'edex 05}\\

and\\

A. Neveu\footnote{neveu@lpmsun2.lpm.univ-montp2.fr. On sabbatical leave
after Sept. 1$^{\hbox{st}}$ 1994 from Laboratoire de Physique Math\'ematique,
Universit\'e de Montpellier II-CNRS, 34095 Montpellier c\'edex 05}

{\em Theoretical Physics Group\\
    Lawrence Berkeley Laboratory\\
      University of California\\
    Berkeley, California 94720}

\end{center}

\vskip .5in

\begin{abstract}
We present numerical evidence that a simple variational improvement of the
ordinary perturbation theory of the quantum anharmonic oscillator can give a
convergent sequence of approximations even in the extreme strong coupling
limit,
the purely anharmonic case. Some of the new techniques of this paper can be
extended to renormalizable field theories.

\end{abstract}
\end{titlepage}
\renewcommand{\thepage}{\roman{page}}
\setcounter{page}{2}
\mbox{ }

\vskip 1in

\begin{center}
{\bf Disclaimer}
\end{center}

\vskip .2in

\begin{scriptsize}
\begin{quotation}
This document was prepared as an account of work sponsored by the United
States Government. While this document is believed to contain correct
 information, neither the United States Government nor any agency
thereof, nor The Regents of the University of California, nor any of their
employees, makes any warranty, express or implied, or assumes any legal
liability or responsibility for the accuracy, completeness, or usefulness
of any information, apparatus, product, or process disclosed, or represents
that its use would not infringe privately owned rights.  Reference herein
to any specific commercial products process, or service by its trade name,
trademark, manufacturer, or otherwise, does not necessarily constitute or
imply its endorsement, recommendation, or favoring by the United States
Government or any agency thereof, or The Regents of the University of
California.  The views and opinions of authors expressed herein do not
necessarily state or reflect those of the United States Government or any
agency thereof, or The Regents of the University of California.
\end{quotation}
\end{scriptsize}

\vskip 2in

\begin{center}
\begin{small}
{\it Lawrence Berkeley Laboratory is an equal opportunity employer.}
\end{small}
\end{center}

\newpage
\renewcommand{\thepage}{\arabic{page}}
\setcounter{page}{1}
\input epsf.tex
\input epsf.sty
\parskip 10pt
\vfill\eject
\setcounter{equation}{0}

\section{Introduction.}

One of us \cite{nev} has proposed a new
approach to non-perturbative phenomena in
 quantum field theory which could combine
 the advantages and range of
 validity of ordinary perturbation theory
 and of variational calculations
of systems with a finite number of degrees
 of freedom. In ref. \cite{nev}, the
 method was applied to the first few orders
 of the anharmonic oscillator whose Lagrangian is:
\begin{equation}L={1\over2}
\left({\partial_{t}\phi}\right)^2
-{1\over2}m^2\phi^2-{1\over4}\lambda
\phi^4\end{equation}

It gave very intriguing results: for example, the
 combination of a very simple variational
 idea with a fifth order perturbative
calculation of the ground-state energy
 at finite $m$ gave in the purely anharmonic
 case $ (m=0)$ a value of the ground state
energy smaller than the true value by only
 $2.10^{-5}$, and use of the seventh
order improved the relative approximation
 to $3.10^{-6}$; similar results were
obtained for the excited states. This
suggests that one could be dealing with
 a convergent sequence of approximations,
 which, if properly generalized to more
complicated cases, could provide a useful
  computational as well as conceptual tool.

In this paper, we present a set of extensive
 numerical investigations on the anharmonic
oscillator treated with this method.
   In spite of the appearance of numbers
 with many significant digits, we do not aim at
finding exact results on the anharmonic
 oscillator, a very well known system.
 On the contrary, this paper should be
 read with a heuristic  point of view
as presenting numerical evidence in
favor of a simple yet potentially
powerful method to variationally
improve perturbation theory in a way
which remains quite empirical. To
 generalize the method to the more
difficult case of a renormalizable
 field theory, one has to understand
how to make it compatible with
 renormalization, and some of the empirical
evidence gathered here turns out to be crucial for
this enterprise. This will be done in other
 publications  \cite{agn}.

Some of the methods used here appear in several
 publications \cite{CAS,ZINN,OKO,GKZ}, under the names
``optimized perturbation theory", ``principle of minimal sensitivity",and
``delta expansion". Here, and in our future publications \cite{agn,bgn},
we provide new insights and expand the range of applications
of these ideas.

 In section 2, exploring perturbation
 theory up to order $n=47$,
 we apply the simple variational procedure of
ref. \cite{nev} to
the calculation of the ground state energy in
 the purely anharmonic case,
 $m=0$, starting from the ordinary perturbative
 expansion at $m\neq0$.
This is equivalent to extrapolating to infinite
 coupling the asymptotic
 expansion in the dimensionless coupling constant
$\lambda/\omega^3$, useful in principle only for
 weak coupling. A rather
 amazing picture emerges; at order
$n$, the procedure yields $n$ values for the
 ground state energy, most of
them complex with very small imaginary parts,
 a few of them real, all of them
within a few percent (most of the time much
much less) from the exact answer,
 $E_0^{exact}$. These values arrange themselves
 in families, increasingly numerous
 as the perturbative order $n$ increases, each
family converging to an approximate
 value of $E_0$, the set of these approximate
values itself having $E_{0}^{exact}$
 as an accumulation point. We give some
arguments to explain this behavior. In
particular, it seems that the variational
procedure generates an effective coupling
 constant which, as the perturbative order
 increases, decreases fast enough to offset
 the well-known factorial increase of the perturbation
theory coefficients. Such an order-dependence of the
effective coupling has been extensively used in refs.
\cite{ZINN,GKZ}, and our findings are within the range of
validity of the results obtained by these authors.
We show how scaling in a natural way the variational
parameter with the order indeed provides much information
about the large order behavior of the procedure, both
for the ordinary anharmonic oscillator and on its
Hartree-Fock approximation which amounts to retaining
only the cactus diagrams: in large order, one obtains a
remarkable improvement of the convergence of perturbation
theory, the variational approximation of the ground-state
energy (say) being then given by a series with  an infinite
radius of convergence. Furthermore, this series can be
computed accurately in perturbation theory, and so the location
of its extrema. Beyond its intrinsic interest, this large
order result is quite important, because it turns out \cite{agn}
that for renormalizable asymptotically free field theories,
the variational procedure is compatible with renormalization
only after the large-order limit has been taken exactly in the
fashion of this section, which thus contains our main results
for future use.

In section 3, we report a study of second
 order of perturbation theory
trying to test the practical usefulness of the
most general variational procedure
 described in ref. \cite{nev}, in
which the propagator is allowed an
arbitrary dependence
on its momentum. We have not been
able to find the exact dependence
which optimizes the second order of
 perturbation theory, but have
tried ansatze with various functional
 forms containing a finite number
of parameters. None of these more
complicated forms seems to improve the
approximation substantially, indicating
 that the simple variational ansatz
of section 2, which modifies the inverse
 propagator by adding a constant,
the result of the lowest order, captures
 most of the physics.
This constant modification is also the
Hartree-Fock approximation,
or large-$N$ if one is dealing with an
$O(N)$ symmetric oscillator.
In section 4, we apply the method to
the calculation of other physical
quantities, like $<\phi^2>$, as a prototype
of the expectation value of
a composite operator, obtaining results
with similar convergence properties.
We also apply the method to the energy
of the first excited state.

\section{A Simple variational
 parameter.}

This section, while self-contained,
is an extension of some
 parts of ref. \cite{nev}.

For the anharmonic oscillator
described by the Lagrangian (1), the
ordinary perturbative
expansion of the ground-state energy
 is of the form:
\begin{equation}
E_{0}^{(p)}(m)={1\over2}m+\sum_{n=1}^{p}
{A_{n}\over4^n}m({\lambda\over m^3})^n,
\end{equation}
where the coefficients $A_n$ can be
found in ref \cite{ben}.
  Apart from the asymptotic behavior of $A_n$
\begin{equation}
A_{n}\sim (-3)^{n}\Gamma(n+1/2)
\end{equation}
which makes the expansion (2) only
Borel summable, this result at finite $p$ is useless as
it stands for $m$ going to zero,
the extreme strong coupling
limit. However, we can introduce
  $\omega$ as a variational
parameter,  rewriting the Lagrangian (1) as:
\begin{equation}
L={1\over2}\left({\partial_{t}\phi}\right)^2
-{1\over2}\omega^2\phi^2-
{1\over2}(m^2-\omega^2)\phi^2-{1\over4}\lambda
\phi^4 = L_{0} + L_{I},
\end{equation}where $L_0={1
\over2}\left({\partial_{t}\phi}\right)^2
-{1\over2}\omega^2\phi^2$ is
our new free Gaussian part.

We can then compute at finite order in
$L_{I}$ the ground state energy. This is
 done easily by starting from
expansion (2), substituting $$m=\omega
\sqrt{1+(m^2-\omega^2)/\omega^2}$$
in it, and expanding
in powers of $m^2-\omega^2$ up to total
order $p$ in $\lambda
$ and $m^2-\omega^2$. Setting $m=0$ then
gives an $\omega$-dependent
approximation of order $p$ for the ground
 state energy of the purely anharmonic case:
\begin{equation}
E_{0}^{(p)}(\omega)=\sum_{n=0}^{p}{A_{n}
\over4^n}{\Gamma(p+n/2+1/2)
\over\Gamma(3n/2+1/2)\Gamma(p-n+1)}
\omega({\lambda\over\omega^3})^n,
\end{equation}
where $A_{0}=1/2$.

  Now, if one would have started from
 the exact expression
for the ground state energy at finite
$m$, the introduction
of $\omega$ should be irrelevant, and
 so, one may take as
best approximation at finite order some
 value of $\omega$
such that $\partial_{\omega}E^{(p)}
(\omega)=0$. This
gives us a set of energy values
which we can compare  with the
numerical value, as
found in ref. \cite{exact}:
\begin{equation}
E_{0}^{exact}(m=0)=\lambda^{1/3}0.420804974478... \,\,.
\end{equation}

 Taking $\lambda=1$ , we have
 performed the calculation
of the optima of the energy in
$\omega$ until the $47^{\rm th}$ order.
 These results and comments on them
 are presented below.

\subsection{Solutions of our
variational problem.}

\begin{figure}
\vspace{-1.5cm}
\centerline{\epsffile{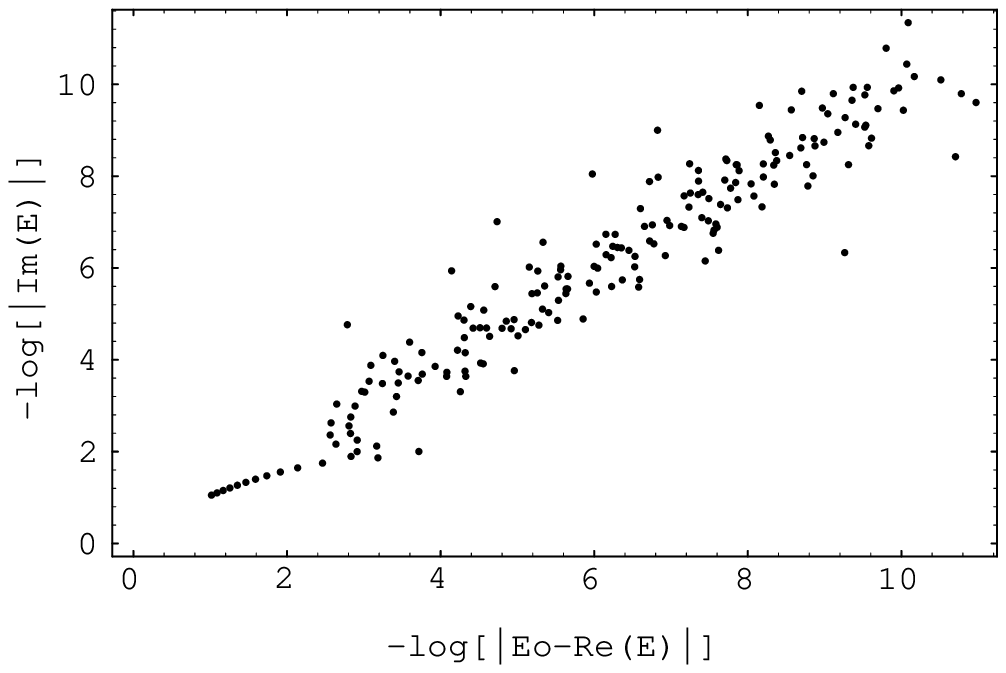}}
\vspace{-1cm}
\caption{In a base-10 log-log diagram, the
imaginary part of the energies
versus the error on their real part.}
\label{log}
\end{figure}

     The search for solutions
 of $\partial_{\omega}E^{(p)}(\omega)=0$
 is performed in the complex plane.
 It gives a polynomial
equation of degree $p$ in $\omega^3 $.
 Most solutions of
this equation are complex and give
complex values for the
 energy, which is of course
unphysical. However, quite remarkably,
 the imaginary part of the energy which one obtains
is generally extremely
small, indeed of the
same order of magnitude
as the error in the real part (see
figure \ref{log}).
 Rather than simply discarding
 these complex extrema,
one may  use them as {\em bona
fide} approximations,
keeping only the real part and
using the imaginary
part as an uncertainty estimate .

   If  the method makes sense, of
adding and subtracting an arbitrary mass term,
 and claiming that if one were working
to all orders this should make no difference,
then, at least in some range of
$\omega$, $E_0^{(p)}(\omega)$
should be less and less dependent on
$\omega$ as
$p$ increases. Indeed, figures
\ref{e0pair} and \ref{e0imp},
where $E_0^{(p)}$ is plotted as
 a function of $\omega$ for
 real $\omega$ and increasing
values of $p$ reveal an increasingly
flat behavior around the best
 value, which provides a way  of
picking the probable best approximation
 when it is real and there
is no imaginary part to serve as an
estimate of the error.

\subsection{Existence of optima families.}

    For a given order $p$,
there are $p$ values of $\omega^3$
which make $E_0^{(p)}$ stationary.
 Figure \ref{om2} shows the distribution
of these values for $\omega^2$ in the first
quadrant of the complex plane,
for all $p$ between 1
and  30 (the reason for choosing
 $\omega^2$ will be clear below).
 Obviously, they arrange themselves
 in families. Closer
examination reveals that as $p$ is
increased by 2,
 one ``new'' complex value appears
 in general,
rather close to the real axis,
which is the
 first member of a new family,
while the other
values obviously are members of
families established  in lower
orders. The new family
generally gives a better
 approximation than the
older ones, which lie
 further away from the
real axis. Once in a while,
instead of one new complex
 family, two new real
values appear. Figure \ref{pvom2}
 plots the values of the real
optimizing values up to order 30.

\subsection{Asymptotic behavior
 of the functions $E_0^{(p)}(\omega)$.}

       We can understand many
of the features presented in
the previous subsection by looking
 at what happens at very high order in eq. (4).

 For any fixed $n$, $p$ very
large, we can use Stirling's formula, and write:
\begin{equation}
E_{0}^{(p)}(\omega)\sim \sum_{n=0}{A_{n}\over4^n}
{1\over\Gamma(3n/2+1/2)}p^{3n/2-1/2}\omega^{1-3n}.
\end{equation}
where we have intentionally left the upper value
of $n$ undefined. Now, from the asymptotic behavior
 (3) $A_{n}\sim (-3)^{n}\Gamma(n+1/2)$
we see that eq. (7) ,
 up to an overall factor $y^{-1/3}$, defines
an entire series in $y=(p/\omega^2)^{3/2})$,
and we call this function
$E_0{(y)}$.

\begin{figure}
\vspace{-1.5cm}
\centerline{\epsffile{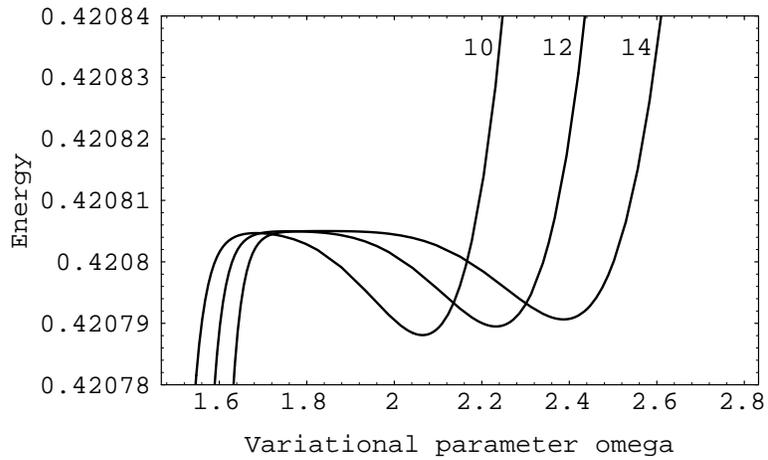}}
\vspace{-1cm}
\caption{$E_{0}(\omega)$ for orders
10, 12 and 14.}
\label{e0pair}
\end{figure}

\begin{figure}
\vspace{-1.5cm}
\centerline{\epsffile{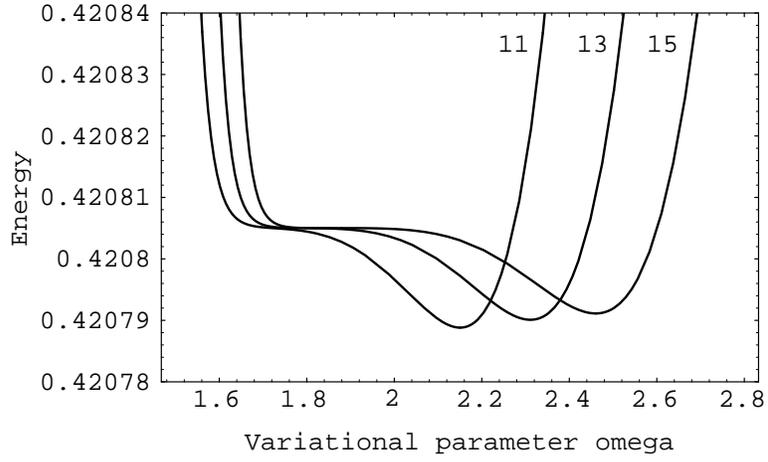}}
\vspace{-1cm}
\caption{$E_{0}(\omega)$ for
orders 11, 13 and 15.}
\label{e0imp}
\end{figure}

\begin{figure}
\vspace{-1.5cm}
\centerline{\epsffile{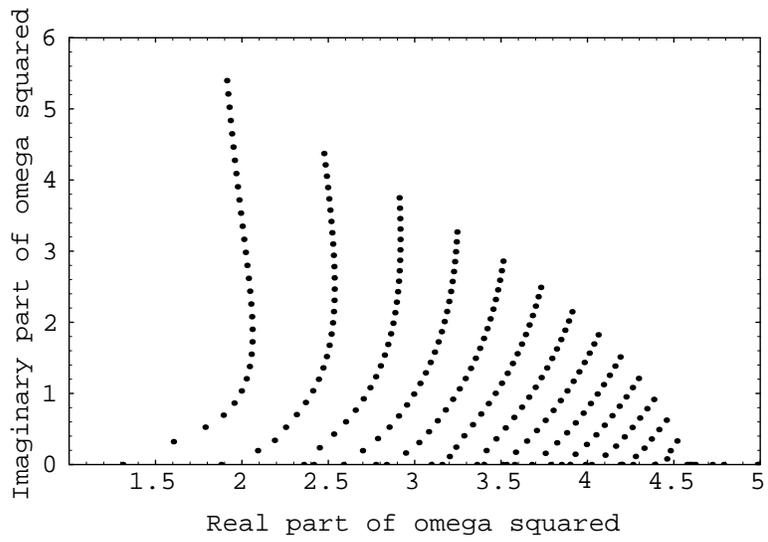}}
\vspace{-1cm}
\caption{Distribution of the
$\omega_{opt}^{2}$ in
the first quadrant of the complex plane.}
\label{om2}
\end{figure}

 Many features of figures \ref{e0pair} and \ref{e0imp}
can be understood in terms of this construction,
which has  manufactured in a
new way an entire series out of an
asymptotic expansion.

  Let us for example explain the
existence of families
 in figures \ref{e0pair} and \ref{e0imp}. When $p$
increases, for finite values
 of $y$, $E_0^{(p)}$ as given
in eq. (4) can increasingly
well be approximated with $E_0(y)$.
 Hence, each extremum of $E_0(y)$
 corresponds to a family.
 Furthermore, we can easily see
that, in each family,
$\omega^2/p$ and the corresponding
 extremum value of $E_0^{(p)}$
 approach their asymptotic value
as $1/p$. Hence, we can fit each
 family with an expansion  in powers
 of $1/p$. Let us do this for
 the first three real families.
For the first one,which starts at
  $6^{\rm th}$ order, we obtain:
\begin{equation}
E_{opt}^{(p)}\simeq 0.4207987 -11.98
\times 10^{-5}/p +8.991
\times 10^{-5}/p^{2} + {\rm O}(p^{-3}) \,.
\end{equation}
For the second one, which begins
 at $15^{\rm th}$ order,
\begin{equation}
E_{opt}^{(p)}\simeq 0.420804977 +5.029
\times 10^{-7}/p -5.284\times 10^{-7}/p^{2}
 + {\rm O}(p^{-3})\,.
\end{equation}
For the third one, which begins
at $30^{\rm th}$ order,
\begin{equation}
E_{opt}^{(p)}\simeq 0.420804974472 -1.25
\times 10^{-9}/p
 -1.38\times 10^{-8}/p^{2} + {\rm O}(p^{-3})\,.
\end{equation}
Including higher orders in $1/p$
does not significantly
change the asymptotic values of these families.

Hence, the first real family forms
 a sequence of approximations which apparently
converges from below to 0.4207986..., which is
significantly, by about $6.10^{-6}$, less than
 the exact value. The second family converges
from above to 0.420804977...,
significantly above the exact value, by only
 $3.10^{-9}$. The third family converges
  to 0.420804974472...,
still significantly below the exact value, by only
about $6.10^{-12}$, which is truly remarkable.
These values correspond to the successive real
extrema of the function $E_0(y)$, which
seems to display an
oscillatory behavior as $y$ increases, which
unfortunately seems to require going to
much higher order than 47
to be seen: at that order, one sees
only the first extremum.
For this extremum, one finds an energy value:
\begin{equation}
E_{0}\simeq 0.42079845\ldots
\end{equation}
Comparing this value with (8), we see
 that it is remarkably
in agreement with the asymptotic value
 of the first family.

It is rather natural that no given
family converge to the
 exact answer, as in practice this
 would essentially
involve only a finite, if extremely
 large, order
of the perturbative series, because
of the convergence
properties of the series in $E_0(y)$.
 The exact value
 can only be the value of $E_0(y)$
at $y=\infty$, as
the data seems to show. So, we can
 try to estimate
the ground state energy by an
extrapolation of $E_0(y)$
 at $y=\infty$.

\subsection{Extrapolation of
 $E_0(y)$.}

Instead of extrapolating $E_0(y)$
to $y=\infty$, we find it more convenient
to perform the change of variable
 $X=y^{-1/3}$ and then extrapolate the function
\begin{equation}
E_{0}(X)=X\sum_{n=0}^{\infty}
{A_{n}\over4^n}{1\over\Gamma(3n/2+1/2)}X^{-3n}
\end{equation}
to $X=0$, knowing that:

i) $E_{0}(X)$ behaves
asymptotically as $X$ for $X\rightarrow \infty$,

  ii) for $X<<1$, this function
has a sequence of very flat extrema, and
goes exponentially fast
to the exact value $E_{0}^{exact}$
as $X\rightarrow \infty$, which is the
number we are looking for.

For this extrapolation, it seems easiest to consider
 the derivative of  $E_0(X)$, which satisfies
\begin{equation}
\lim_{X\rightarrow\infty}\partial_{X}
E_{0}(X)={1\over{2\sqrt{\pi}}}\, ,
\end{equation}
 impose  a decreasing
exponential behavior at
 $X=0$, and find the value of $E_0(X=0)$ by
integration from infinity to zero.
For example, we can fit
 this function, $\partial_{X}
E_{0}(X)$, at $X\rightarrow\infty$,
using the functional form

\begin{equation}
\partial_{X}E_{0}(X)=
{1\over{2\sqrt{\pi}}}\exp(-a/X^3)
[1+b/X^3+c/X^6+\ldots].
\end{equation}
with appropriate values of the parameters $a,b,c\ldots$

This extrapolation procedure can be considered as
an alternative to the extremization
procedure of the previous subsections,
using the same numerical information,
encoded in the values of the successive
perturbation theory coefficients
$A_0,A_1,A_2\ldots$ We thus obtain:

  - Using only $A_0$ and $A_1$, $i.e.$
setting $b=c=\ldots =0$:
\begin{equation}
E_{0}^{appr}=0.420014\ldots
\end{equation}which must be
compared with the usual variational
 value $E_{0}^{(1)}=0.429268\ldots$
 and the exact
 $E_{0}^{exact}\simeq 0.420805$.
 Remarkably, taking into account only
the first order
 Feynman diagram, we thus obtain an
approximate value
 below the exact ground state energy
 by only $8\times10^{-3} $! This is actually
due to the fact that in the next
approximations, the coefficient $b$
is accidentally very small. An accuracy
of a few percent would probably be more generic.

  - Using $A_0$, $A_1$ and $A_3$,
\begin{equation}
E_{0}^{appr}=0.4204619\ldots
\end{equation}
to be compared with $E_{0}^{(2)}
=0.421218\ldots +0.0014\ldots {\rm i}$,

  - Using $A_0$, $A_1$, $A_3$ and $A_4$,
\begin{equation}
E_{0}^{appr}=0.420474\ldots
\end{equation}
to be compared with $E_{0}^{(3)}=0.420983\ldots$

We see that this extrapolation method of the
entire function $E_0(y)$ gives remarkably
good numerical values, and has the further
advantage of giving {\em real} numbers. The
functions analogous to $E_0(y)$ will play
a crucial r\^{o}le in the quantum field
theory case, as it turns out to be the
main feature of this paper which survives
renormalization and its infinities.

\subsection{Oscillatory behavior
 of $E_{0}(X)$ in the case
 of the Hartree-Fock approximation.}

In this subsection, we report the results
of the procedures of the previous
subsections applied to
the theory restricted to the Hartree-Fock
approximation (if we were dealing with
an $N$-component symmetric oscillator, this
would correspond to taking the large-$N$ limit).
The advantage of the Hartree-Fock case
is that ordinary perturbation theory
of the Lagrangian (1) has a finite radius
of convergence in $\lambda /m^3$. We
shall see that  the features
discovered in the previous subsections
are naturally also present in this
approximation, together with a couple
of extra ones.

The Hartree-Fock approximation of the
 ground-state energy
 is given by the sum of all vacuum to
 vacuum cactus Feynman
diagrams. Applying the same procedure
 as in section 1, we obtain for the order-$p$
approximation at $\lambda =1$:
\begin{equation}
E_{H.F.}^{(p)}(\omega)=\sum_{n=0}^{p}
{C_{n}\over4^n}{\Gamma(p+n/2+1/2)
\over\Gamma(3n/2+1/2)\Gamma(p-n+1)}
\omega({1\over\omega^3})^n \,,
\end{equation}
where the coefficients $C_{n}$ give
 the sum of all cactus graphs
of order $n$. This is the expression,
identical in form to equation (5),
which has to be studied in our variational
approximation.

\begin{figure}
\vspace{-1.5cm}
\centerline
{\epsffile{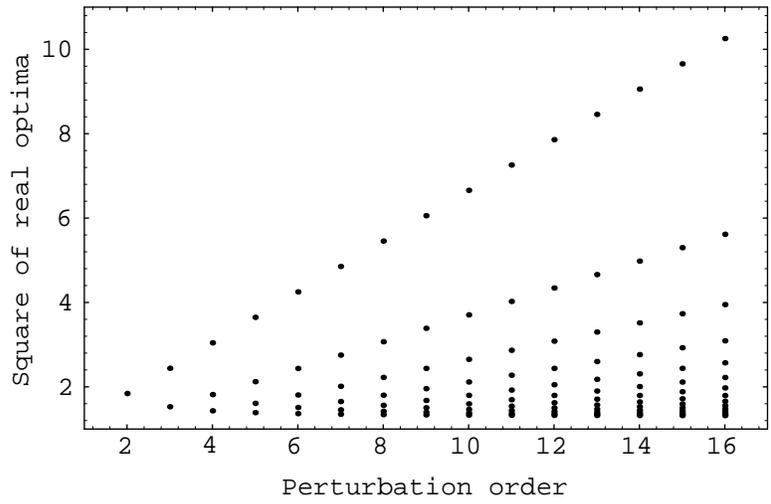}}
\vspace{-1cm}
\caption{The real $\omega_{opt}^{2}$
versus the
perturbation order in the Hartree-Fock
 approximation.}
\label{hfpom2}
\end{figure}

\begin{figure}
\vspace{-1.5cm}
\centerline{\epsffile{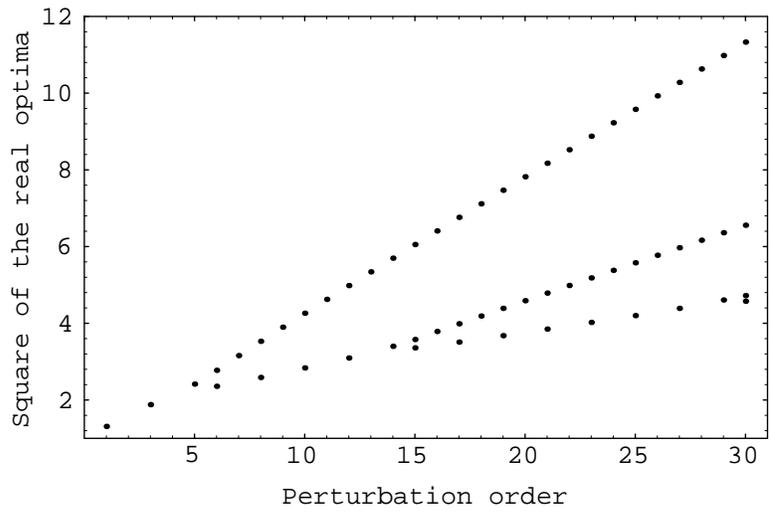}}
\vspace{-1cm}
\caption{The  real $\omega_{opt}^{2}$ versus
 the perturbation order.}
\label{pvom2}
\end{figure}

In the framework of this approximation,
 we obtain many real
 solutions of the variational equation
(see figure \ref{hfpom2}) to be compared
 with the exact value

\begin{equation}
E_{H.F.}^{exact}=0.4292678409575 \ldots
\end{equation}

Remarkably, at any order, the exact value
is one of the solutions of the variational
procedure, the one corresponding to the smallest
value of $\omega$, which sits precisely
at the edge of the convergence
region of the original perturbative series.
This coincidence is of course a
pathology unique to the Hartree-Fock
approximation, which corresponds to
the fact that in the full theory, the
variational value for the energy which
gives the best numerical value is the one
for the smallest $\omega$. Apart from that,
the grouping of the values of $\omega$ in
families, and the oscillatory behavior
of the function $E_{H.F.}^{(p)}(\omega)$ are much
the same as in the full theory.

Next, as in subsection (2.3), we go to the limit
$p\rightarrow\infty$ with the same rescaling
of $\omega$, and define the function:

\begin{equation}
E_{H.F.}(y)=y^{-1/3}(\sum_{n=0}^{\infty}
{C_{n}\over4^n}{1\over\Gamma(3n/2+1/2)}y^n).
\end{equation}

\begin{figure}
\vspace{-1.5cm}
\centerline{\epsffile{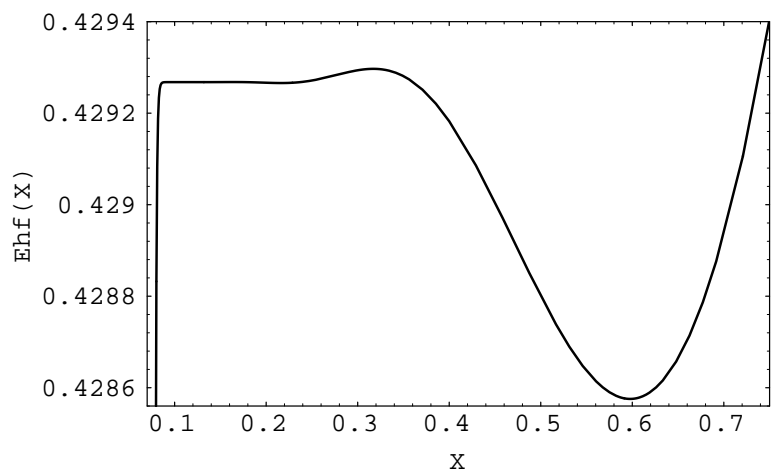}}
\vspace{-1cm}
\caption{$E_{H.F.}(X)$ where $X=y^{-1/3}$,
plotted using the
 50 first terms of the series (20).}
\label{e0hf}
\end{figure}

\begin{figure}
\vspace{-2.5cm}
\centerline{\epsffile{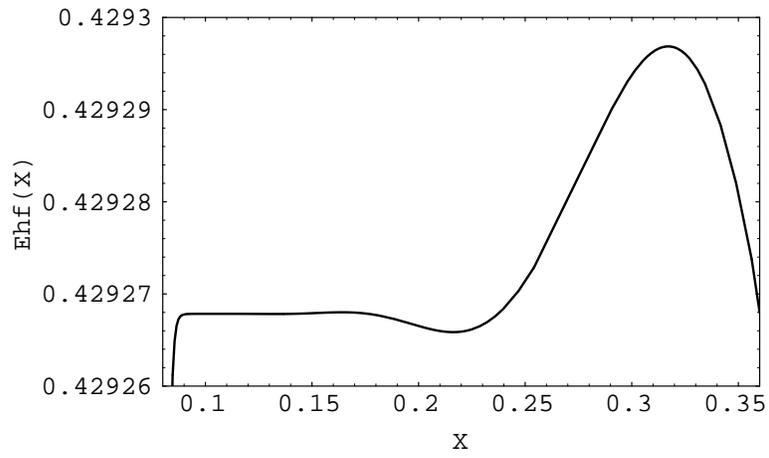}}
\vspace{-2.5cm}
\vspace{-1cm}
\centerline{\epsffile{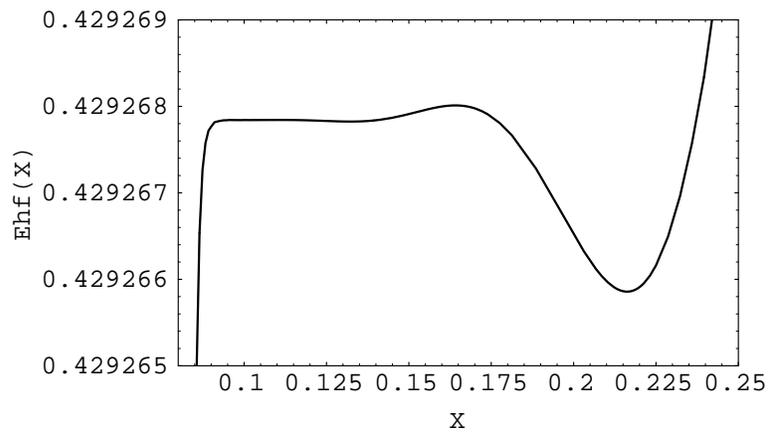}}
\vspace{-1cm}
\caption{Zooms of figure 7.}
\label{e0hfgp}
\end{figure}

Because the $C_n$ coefficients are those of
a series with finite radius of convergence,
one can obtain much better estimates of the
series in this equation for large values of $y$,
and thus see very well its oscillating
behavior in that region, which was not the case
for the full theory. This is displayed in figures
\ref{e0hf} and \ref{e0hfgp}.

We can for example compare several asymptotic energy
values associated with the families displayed
in fig. \ref{hfpom2} with the stationary values
of $E_{H.F.}(y)$. For the first three
families,we find:

-\underline{ First family}:
\begin{equation}
E_{opt}^{(p)}= 0.428584 - 0.00382794/p +
 0.00244357/p^2 + {\rm O}(p^{-3}),
\end{equation}
and for the corresponding extremum of $E_{H.F.}(y)$:
\begin{equation}
E_{H.F.}^{ext}=0.42857\ldots
\end{equation}

-\underline{ Second family}:
\begin{equation}
E_{opt}^{(p)}= 0.4292937645 - 0.00051/p
+ 0.00031/p^2 + {\rm O}(p^{-3}),
\end{equation}
and
\begin{equation}
E_{H.F.}^{ext}=0.429297\ldots
\end{equation}

-\underline{ Third family}:
\begin{equation}
E_{opt}^{(p)}= 0.4292674287 - 0.000096/p
 -0.00021/p^2 + {\rm O}(p^{-3}),
\end{equation}
and
\begin{equation}
E_{H.F.}^{ext}=0.4292659\ldots
\end{equation}

By applying our variational procedure to
the Hartree-Fock approximation and comparing
with the full theory, we thus see which features
are more generic, and hence have a better chance
of surviving in field theory. We think that the
main lesson to be learned is that the exact
Hartree-Fock result, already found at lowest order,
and common in that order to the Hartree-Fock case
 and to the complete theory
may be misleading, in the sense that it seems
to give too much weight to the fact that the
variational method gives the exact result for
the Hartree-Fock case. In contrast to this
possibly pathological coincidence, the behavior
of the large-order limit function $E_0(X)$ is
common to the Hartree-Fock case and to the
complete theory. This large-order limit also sheds
light on the general behavior of the curves of figures
\ref{e0pair} and \ref{e0imp} : the regular shift to the
right of the right-hand parts of the curves exactly corresponds
to the rescaling of $\omega^2$ by the order in the large-order
limit. For the left-hand parts of the curves, they do not shift to
the right in the case of the Hartree-Fock approximation, and
shift to the right in the complete theory only as $\omega^3$
proportionnal to the order. Thus, in both cases, there is an increasing
 region in $\omega$ where the curves become flatter as the order increases
and hence, the approximation more and more accurate. These scaling
behaviors correspond to those found in reference \cite{GKZ}.
Furthermore, the limiting  function $E_0(X)$ can
be computed in perturbation theory for large
$X$, and, as we have seen in subsection (2.4),
accurately extrapolated to $X=0$ to give an excellent
value of the energy, and we shall see in our
extension to quantum field theory \cite{agn}
that this is indeed
the most robust feature of this section, robust
enough to survive renormalization.

  \section{More sophisticated approach:
the use of several parameters.}

   In the previous section, our variational improvement
 of perturbation theory has been severely restricted
to  the class of variational ansatze where the modification
 of the inverse propagator is a constant. It would be
worthwhile to have an idea of the improvement which a
more general modification would give.

\begin{figure}
\centerline{\epsffile{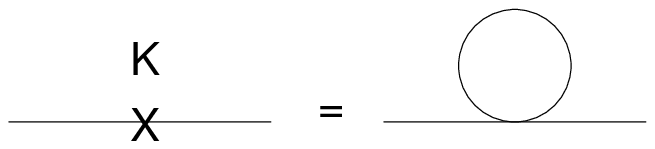}}
\caption{First order ``gap
equation".}
\label{gap1}
\end{figure}

\begin{figure}
\centerline{\epsffile{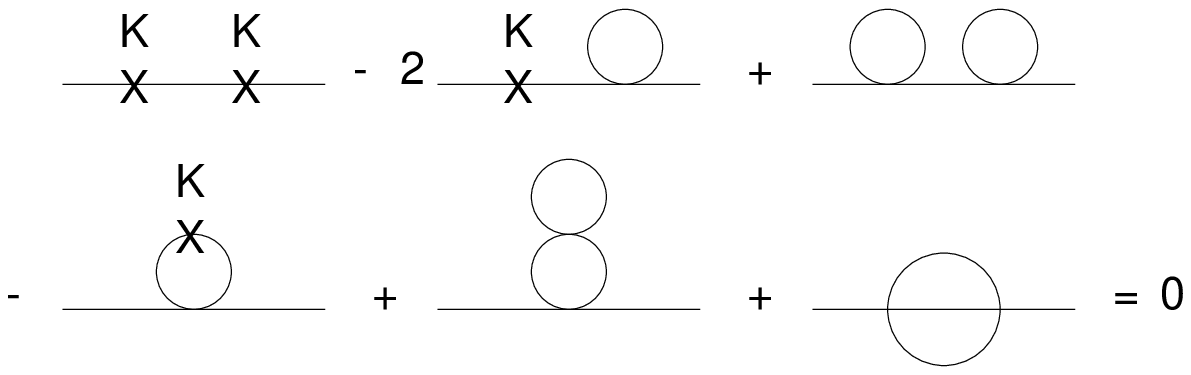}}
\caption{Second order ``gap
equation".}
\label{gap2}
\end{figure}

  The variational improvement of perturbation
theory provides  an order-dependent equation for the
modified propagator
$G=1/(k^2-K)$ which at first
and second orders is depicted by the Feynman diagrams
 of figures \ref{gap1} and \ref{gap2}, in which the
 internal lines involve the modified propagator $G$
itself.

In principle, the most arbitrary $G$ could
involve a complicated kernel, but it is obvious that
 there should exist solutions of these equations
of the form $G(k)=1/[k^2-K(k)]$
(more general kernels $G(k,k')$ would occur naturally
if one were looking for solutions corresponding
to bound states, and must exist,
as indeed the functional equation
of figure \ref{gap1} is identical to equation (2.14)
of reference \cite{DHN} which has a very rich
set of solutions. However, pursuing
investigations in that direction would go
beyond the scope of the present paper).
 In lowest order,
 a solution of this form will automatically lead
 to a constant $K$, solution of the Hartree-Fock
 approximation. In second order, because of the
non-trivial structure of the last diagram in
 Fig. \ref{gap2},
$K(k)$  must be non constant. The equation which $K$
satisfies is an unusual non-linear integral equation
for which we have found no clever trick. Hence, we
resort once more to a further approximation, in which
we restrict $K(k)$ to certain functional forms with
a finite number of variational parameters in them
with respect to which we optimize. We now list these
 various functional forms in Euclidean notation and
the corresponding results of the optimization.

     For reference, we  first give the values for the simple
 ansatz of the previous section:

 i)For $G(k)=1/(k^2+m^2)$ :
\begin{equation}m\simeq 1.2980+0.0083\, {\rm i}\end{equation}
which gives:
\begin{equation}E_{opt}\simeq 0.42142+0,00013\,{\rm i}\end{equation}

  Moving to increasingly complicated ansatze, we have:

     ii)For $G(k)=\alpha /(k^2+m^2)$ :\begin{eqnarray}
\alpha&\simeq& 0.9836+0.1511\,{\rm i}
\\m&\simeq& 1.3102+0.0427\,{\rm i} \nonumber
\end{eqnarray}
which give:
\begin{equation}E_{opt}\simeq 0.42169+0.00014\,{\rm i}\,.
\end{equation}
and
\begin{eqnarray}\alpha&\simeq& 0.9423+0.087\,{\rm i}\\
m&\simeq& 1.1957+0.1915\,{\rm i} \nonumber
\end{eqnarray}which give:
\begin{equation}E_{opt}\simeq 0.42049+0.00152\,{\rm i}\,.
\end{equation}

    iii)For $G(k)=\alpha /(k^2+m^2)+(1-\alpha)/(k^2+n^2)$ :
\begin{eqnarray}\alpha&\simeq& 0.8979-0.0442\,{\rm i} \nonumber \\
m&\simeq& 1.1833+0.0655\,{\rm i}
\\n&\simeq& 7.6246-2.0266\,{\rm i} \nonumber
\end{eqnarray}which give:\begin{equation}
E_{opt}\simeq 0.42096+0.00099\, {\rm i}\,.\end{equation}

iv)For $G(k)=\alpha /(k^2+m^2)+\beta/(k^2+n^2)
+(1-\alpha-\beta)/(k^2+u^2)$ :
\begin{eqnarray}
\alpha &\simeq& 0.9105+0.0678\,{\rm i} \nonumber \\
\beta &\simeq& 0.0989-0.0233\,{\rm i} \nonumber \\
m &\simeq& 1.1902-0.0458\,{\rm i}\\
n &\simeq& 3.1798+5.1317\,{\rm i} \nonumber \\
u &\simeq& 5.1534+18.30\,{\rm i} \nonumber
\end{eqnarray}
which give:
\begin{equation}
E_{opt}\simeq 0.42095-0.00098\, {\rm i}\,.
\end{equation}
    iv)For $G(k)=1/(k^2+\alpha |k|+m^2)$ :
\begin{eqnarray}
\alpha &\simeq& 0.1751-0.1459\,{\rm i}\\
m &\simeq& 1.2222+0.1509\,{\rm i} \nonumber
\end{eqnarray}
which give:
\begin{equation}
E_{opt}\simeq 0.42084+0.00129\,{\rm i}\,.
\end{equation}
and
\begin{eqnarray}
\alpha &\simeq& 0.0198+0.2957\,{\rm  i}\\
m &\simeq& 1.2844+0.0459\,{\rm i} \nonumber
\end{eqnarray}
which give:
\begin{equation}
E_{opt}\simeq 0.42149+0.00090\,{\rm i}\,.
\end{equation}

We should note that these results are not as exhaustive as in
the previous section, because we must look for the
extrema in the complex planes of several parameters,
 where it is easy to get lost: there may be other
extrema, in some far away regions of parameter space.
 Nevertheless, we believe that the following
qualitative picture emerges:
Such optimizations do not seem to increase
 appreciably the accuracy on the ground state energy whether
 measured by the value of
the imaginary part of $E_{opt}$ or by the difference of the
real part with the true value: this accuracy remains around
$10^{-3}$. We may reasonably conjecture that
 this is indeed as well as the method
can do at this second order, and that solving
the variational equation
 exactly for the propagator at that order would not give any
substantial improvement beyond the result of the
simple ansatz of the previous section.

   \section{Calculation of other physical quantities.}
\subsection{Mean value of $<\phi^{2}>$:}

     Vacuum expectation values of composite
 operators are objects
 of great physical interest in quantum
field theories, much more than in ordinary
quantum mechanics. For example, they play
an important r\^{o}le in chiral perturbation
theory, and it would be most welcome to have
even rather crude approximations for their values.
Hence, in this section, we shall investigate
the accuracy with which our method  can give
 $<\phi^2 >$ in the ground state of the
anharmonic oscillator.

We shall also use this calculation to test another
idea: although it is
clear  that, as in any variational procedure,
 one would in principle get the best results by
optimizing the quantity of interest with respect to
the variational parameters, it is sometimes
tempting to use the values  of the
parameters obtained in the
optimization of some quantity (in the present case
the ground state energy) in order to compute
other quantities with them, particularly
if these other quantities give rise to some
unwieldy set of variational equations (this could
occur for example for a propagator
or some more complicated correlation function
which have some space-time variables in them).
We are thus going to compare the results of these
two procedures for the ground
 state expectation value
$<\phi^2>$. From
\begin{equation}
E_0(m,\lambda)=<-{1\over2}{d^2\over d\phi^2}
+{1\over2} m^2 \phi^2+\frac{\lambda}{4} \phi^4>,
\end{equation}
we have simply
\begin{equation}
<\phi^2>=2\frac{\partial}{\partial m^2} E_0(m,\lambda)
\end{equation}
so that the  perturbative expansion at non-zero $m$
of $<\phi^2>$ is rather  trivially obtained
from equation (2), and one can then proceed
as in section (2):
  At $m=0$, we have thus computed the
 optima with respect to $\omega$
up to order 30, and compared them with the value
found in ref. \cite{baner}:
\begin{equation}<\phi^2>_{exact}=
\lambda^{-1/3}\,\, 0.456119955748\ldots
\end{equation}

We have found  convergence
and accuracy properties very similar
 to those of the ground-state energy
 itself as described in section (2), and we
do not report here the details.

Next, we compare the results of optimizing
directly $<\phi^2>$ with the value obtained
by using the $\omega$ obtained from the
optimization of the ground-state energy in the
same order. For the direct calculation to
orders $p=1$ to 3, we obtain:
\begin{eqnarray}
{\rm for }\,\,\,p=1:\,\, &  \omega \simeq 1.25992
  &    <\phi^2> \simeq 0.44645 \nonumber \\
\nonumber \\ {\rm for }\,\,\,p=2:\,\,
 &\,  \omega \simeq 1.34807 + 0.11106\,{\rm i}\, &\,
<\phi^2> \simeq 0.45575 - 0.00227\,{\rm i}
\nonumber \\ \\ {\rm for }\,\,\,p=3:\,\,
 &  \omega \simeq 1.43806 &    <\phi^2> \simeq 0.45592
 \nonumber \\ &  \omega \simeq 1.40288 + 0.17821\,{\rm i}\,
 &\,    <\phi^2> = 0.45723 - 2.7\times
 10^{-5}{\rm i} \nonumber \end{eqnarray}
while using the values of $\omega$
obtained from the optimization of the
ground-state energy at the same order, we obtain
\begin{eqnarray}
{\rm for }\,\,\,p=1:\,\,
 &  \omega \simeq 1.14471  &
<\phi^2> \simeq 0.43678 \nonumber \\
\nonumber \\ {\rm for }\,\,\,p=2:\,\,
 &\,  \omega \simeq 1.27264 + 0.12645 \, {\rm i}\, &\,
<\phi^2> \simeq 0.45424 - 0.00389\,{\rm i}\,
\nonumber \\ \\ {\rm for }\,\,\,p=3:\,\,
 &  \omega \simeq 1.37080 &    <\phi^2> \simeq 0.45549
\nonumber \\ &  \omega \simeq 1.35091 + 0.19488\,{\rm i}\,
 &\,    <\phi^2> \simeq 0.45744 - 6.8\times 10^{-4}
{\rm i} \nonumber \end{eqnarray}

  Comparing these numbers with
the exact value (43), we see
that these two approaches lead to results with
 the  same order of accuracy, the direct
optimization doing only slightly better.
In the next subsection, we shall perform a similar
comparison for another quantity.

   \subsection{Energy of the first excited state }
      We can use the perturbative expansion
for the energy of the first
excited state,
\begin{equation}
E_1(m,\lambda)=\frac{3}{2} m+\frac{15}{16}
\frac{\lambda}{m^2}-\frac{165}{128}\frac{\lambda^2}{m^5}
+\frac{3915}{1024}\frac{\lambda^3}{m^8}
\end{equation}
and repeat on it the optimization of section (2)
up to order 3. We obtain a sequence of
approximations with very similar features. The
results for the first three orders $p$ are:
\begin{eqnarray}
{\rm for }\,\,\,p=1:\,\,
 &  \omega \simeq 1.35721  &    E_{1}^{(1)} \simeq 1.52686 \\
\nonumber \\ {\rm for }\,\,\,p=2:\,\, &\,
 \omega \simeq 1.49893 + 0.08752\,{\rm i}\, & \,   E_{1}^{(2)} \simeq
 1.50739 + 0.001\,{\rm i} \nonumber \\ \nonumber
 \nonumber \nonumber \nonumber \nonumber \nonumber \\
{\rm for }\,\,\,p=3:\,\, &  \omega = 1.69418 &
  E_{1}^{(3)} \simeq 1.50710 \nonumber \\
 &  \omega \simeq 1.53709-0.11684\,{\rm i}\,
  &\,    E_{1}^{(3)} \simeq 1.50750+5.8\times
 10^{-4}{\rm i} \nonumber\end{eqnarray}
which can be compared to the accurate value:
\begin{equation}
E_{1}^{exact}=1.50790125\ldots
\end{equation}

One may now compare these results with
those obtained by evaluating $E_1^{p}$ at the optima of
$E_0^{p}$ for  the same value of $p$.
This gives:
\begin{eqnarray}
{\rm for }\,\,\,p=1:\,\,
 &  \omega \simeq 1.14471  &    E_{1}^{(1)} \simeq 1.57398 \\
\nonumber \\ {\rm for }\,\,\,p=2:\,\, &\,
 \omega \simeq 1.29802 + 0.00828\,{\rm i}\, &\, E_{1}^{(2)}\simeq
 1.49323 + 0.0016\,{\rm i} \nonumber \\ \nonumber
 \nonumber \nonumber \nonumber \nonumber \nonumber \\
{\rm for }\,\,\,p=3:\,\, &  \omega \simeq 1.37080 &
  E_{1}^{(3)} \simeq 1.51382 \nonumber \\
 &  \omega \simeq 1.35091-0.19449\,{\rm i}\,
  & \, E_{1}^{(3)} \simeq 1.49385+0.00075\,{\rm i}
\nonumber\end{eqnarray}

Although the true optima of $E_1^{(p)}$ are
indeed much closer to the exact answer than
the values at the  optima of $E_0^{(p)}$, these
nevertheless provide quite reasonable
approximations, which improve when the order
increases, due to the increasingly flat character
of the corresponding function, analogous to
what can be seen on figures 2 and 3.

\begin{figure}
\centerline{\epsffile{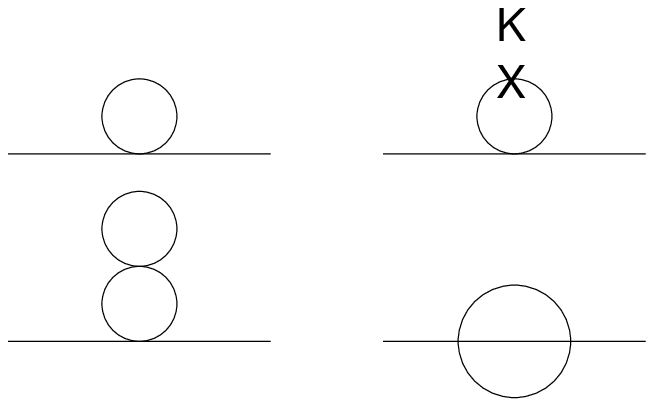}}
\caption{1-P-I diagrams of
 $\Gamma^{(2)}(k)$.}
\label{1pi}
\end{figure}

Another way to compute the energy of the
first excited state is to use the 2-point 1-P-I
function at order $p$, $\Gamma^{(p)}(k)$ which vanishes
for $k=E_{1}-E_{0}$. Figure \ref{1pi} shows
 the Feynman diagrams
to be computed up to second order of
perturbation theory.

In general, a direct optimization of this
function would have to be made for each value
of $k$. In first order, taking $1/(k^2+\omega^2)$
as variational ansatz, one obtains
\begin{equation}
\Gamma_{(2)}(k,\omega)=k^2+\frac{3}{2}\frac{\lambda}
{\omega},
\end{equation}
which has no extremum in $\omega$: the only natural
thing to do at that order is thus to plug in the
value of $\omega$ obtained in the optimization of
the ground-state energy: this gives back the Hartee-Fock
value for the mass gap, $1.1447142\ldots$
(at $\lambda=1$), while the correct value is
$E_1^{exact}-E_0^{exact}=1.087096\ldots$

In second order, taking again simply $1/(k^2+\omega^2)$
as variational ansatz, one obtains:
\begin{equation}
\Gamma_{(2)}(k,\omega)=k^2-\frac{9}{4\omega}
+\frac{9\lambda}{2\omega}\bigl(\frac{1}{4\omega^2}
+\frac{1}{k^2+9\omega^2}\bigr).
\end{equation}

Hence, one can solve the sytem
\begin{eqnarray}
\partial_{\omega}\Gamma_{(p)}(k,\omega)=0,\\
 \Gamma_{(p)}(k,\omega)=0,
\end{eqnarray}
obtaining at $\lambda=1$:
\begin{eqnarray}
\omega\simeq 1.4107698 \\
k=E_1-E_0\simeq1.092026
\end{eqnarray}
which is about $10^{-3}$ away from the correct value,
{\em i.e.} an error of the same order of magnitude
as in the calculations of $E_0$ or $E_1$ at the
same order, except that it has the advantage
of being real (by chance, as far as we can tell).
By contrast, plugging in $\Gamma_{(p)}(k,\omega)$
the (complex) value of $\omega$ obtained from
the optimization of $E_0^{(2)}$ and looking
for the zero gives an error of a few percents
in $E_1-E_0$, compared to the $\simeq 10^{-3}$
error seen in equation (55). Using the more
complicated anstze of the end of section 3
give similar results, and we do not report
them here.

In this section, we have seen that the
procedure of \cite{nev} can be used successfully
to compute a variety of physical quantities.
In each case, as expected from a variational
approximation, directly optimizing the
quantity of interest gives the best answer, but
using the values of the variational parameters
from the optimization of another quantity
is not necessarily bad: the loss of numerical
accuracy  seems  of the same order as working
in one order less in perturbation theory.

\section{Conclusion}

In this first paper, we have first explored in numerical
detail the remarkable convergence properties of the
variational method proposed in \cite{nev}
on the particular case of the ground-state
energy of the anharmonic oscillator. We have found
how the extrema arrange themselves in families,
and how these families converge to the exact answer.
We have seen that these families are also present
in the Hartree-Fock approximation, and in the
calculation of other physical quantities.

Although  higher orders of the variational
procedure do not give bounds (contrary to the usual
lowest order), this is compensated by its
straighforward applicability to the calculation
of excited states or expectation values.

Going to infinite order in the
(easy) variational procedure (which involves only
a modification of the free Lagrangian)
but finite order
in the (hard) perturbative calculation
(which involves the quartic interaction term),
we have put some order in these families,
 improved the convergence of the perturbative
series by a factorial, and defined an extrapolation
procedure which is not only numerically excellent,
but will be seen in  future publications \cite{agn} to
generalize to the case of renormalizable field theory.

In the next paper \cite{bgn}, we shall see how a similar
variational improvement of perturbation theory can be
achieved through a variational ansatz corresponding
to putting the anharmonic oscillator in a finite time box.

\section{Acknowlegments}

One of us (A. N.) is grateful for their
hospitality to the  Lawrence Berkeley Laboratory,
where this work was supported in part by the Director,
Office of Energy research, Office of High Energy and
Nuclear Physics, Division of High Energy Physics of
the U.S. Department of Energy under Contract
DE-AC03-76SF00098 and to the Physics Department,
University of California,
Berkeley, where this work was completed with partial
support from National Science Foundation
grant PHY-90-21139.

\vfill
\newpage

\begin{thebibliography}{aaaaaaa}

\bibitem{nev}
A.Neveu, Nucl. Phys. B 18 (1990) 242.

\bibitem{agn}
C. Arvanitis, F. Geniet and A. Neveu,
``Variational solution of the Gross-Neveu model,
I. The large-$N$ limit", Montpellier preprint PM94-19
(hep-th/9506188);\\
C. Arvanitis, F. Geniet, J.-L. Kneur, M. Iacomi and
A. Neveu, ``Variational solution of the Gross-Neveu model,
II. Finite-$N$ and renormalization", Montpellier preprint PM94-20,
 to be published.

\bibitem{CAS}
W.E. Caswell, Ann. Phys. (N.Y) 123 (1979) 153;\\
I.G.  Halliday and P. Suranyi, Phys. Lett.
 85 B (1979) 421; \\
P.M.  Stevenson, Phys. Rev.  D 23 (1981) 2916;\\
I. Stancu and P.M.  Stevenson, Phys. Rev.
 D 42 (1990) 2710;\\
J. Killinbeck, J. Phys.  A 14 (1981) 1005.

\bibitem{ZINN}
R. Seznec and J. Zinn-Justin, J. Math.
 Phys. 20 (1979) 1398.

\bibitem{OKO}
A. Okopinska, Phys. Rev.  D 35 (1987) 1835;\\
A. Duncan and M. Moshe, Phys. Lett.
 215 B (1988) 352; \\
H.F. Jones and M. Moshe, Phys. Lett.
 234 B (1990) 492;\\
S. Gandhi, H.F. Jones and M. Pinto, Nucl. Phys.
 B 359 (1991) 429; \\
S. Gandhi, M. Pinto, Phys. Rev. D 46 (1992) 2570;\\
A.N. Sissakian, I.L. Solovtsov and O.P. Solovtsova
Phys. Lett. 321 B (1994) 381;\\
A. Duncan and H.F. Jones, Phys. Rev. D 47 (1993) 2560; \\
C.M. Bender, A. Duncan and H.F. Jones,
Phys. Rev. D 49 (1994) 4219; \\
 C. Arvanitis, H.F. Jones and C.
Parker, Imperial College preprint TP/94-95/20
(hep-ph/9502386) to be published.

\bibitem{GKZ}
R. Guida, K. Konishi and H. Suzuki, Genova
Preprints  GEF-Th-7/1994
 (hep-th/9407027) and  GEF-Th-4/1995
 (hep-th/9505084).

\bibitem{ben}
C.M. Bender and T.T. Wu, Phys. Rev. 184 (1969) 1231.

\bibitem{exact}
C.M. Bender, K. Olaussen and P.S. Wang,
Phys. Rev. D 16 (1977) 1780.

\bibitem{baner}
K. Banerjee, S.P. Bhatnagar, V. Choudhry and S.S. Kanwal,
 Proc. R. Soc. Lond. A 360 (1978) 575-586.

\bibitem{bgn}B. Bellet, P. Garcia and A. Neveu,
``Convergent sequences of perturbative approximations
for the anharmonic oscillator, II. Compact time approach",
Montpellier preprint, next paper.

\bibitem{DHN}
R.F. Dashen, B. Hasslacher and A. Neveu, Phys. Rev.
 D 12 (1975) 2443.
\end{thebibliography}
\end{document}